\documentclass{svjour2}                    
\smartqed  
\usepackage{graphicx}
\usepackage{braket}
\usepackage{bbm}
\usepackage{amssymb}
\usepackage{amsmath}
\usepackage{mathptmx}      
\newcommand{\mean}[1]{{\langle #1 \rangle}}
\newcommand{\BRA}[1]{{\langle #1 |}}
\newcommand{\KET}[1]{{| #1 \rangle}}
\newcommand{\BRAKET}[2]{{\langle {#1} | {#2} \rangle}}
\newcommand{\bm}[1]{\boldsymbol{#1}}
\newcommand{\bmsigma}{{\bm{\sigma}}}

\newcommand{\GI}{{ G}}
\newcommand{\caI}{{\mathbb I}}
\newcommand{\caV}{{\mathbb V}}
\newcommand{\bbR}{{\mathbb R}}
\newcommand{\caL}{{\mathcal L}}
\newcommand{\caR}{{\mathcal R}}
\newcommand{\id}{\textrm{d}}

\DeclareMathOperator{\Prob}{Prob}
\DeclareMathOperator{\Tr}{Tr}

\begin{document}

\title{Computation of current cumulants for small nonequilibrium systems}
\author{ Marco Baiesi \and Christian Maes \and Karel {Neto\v{c}n\'{y}} }


\institute{Marco Baiesi \at
           Instituut voor Theoretische Fysica, K.~U.~Leuven,
B-3001 Leuven, Belgium \\
           \email{marco@itf.fys.kuleuven.be}
           \and
           Christian Maes \at
           Instituut voor Theoretische Fysica, K.~U.~Leuven,
B-3001 Leuven, Belgium \\
           \email{christian.maes@fys.kuleuven.be}
           \and
           Karel Neto\v{c}n\'{y} \at
           Institute of Physics AS CR, Prague, Czech Republic.\\
           \email{netocny@fzu.cz}
}

\date{Received: date / Accepted: date}

\maketitle

\begin{abstract}
We analyze a systematic algorithm for the exact computation of the
current cumulants in stochastic nonequilibrium systems, recently
discussed in the framework of full counting statistics for
mesoscopic  systems. This method is based on identifying the
current cumulants from a Rayleigh-Schr\"odinger perturbation
expansion for the generating function. Here it is derived from a
simple path-distribution identity and extended to the joint
statistics of multiple currents. For a possible thermodynamical
interpretation, we compare this approach to a generalized
Onsager-Machlup formalism. We present calculations for a
boundary driven Kawasaki dynamics on a one-dimensional chain, both
for attractive and repulsive particle interactions.

\keywords{current fluctuations \and nonequilibrium \and cumulant
expansion} \PACS{05.70.Ln \and 05.10.Gg \and 05.40.-a \and
82.60.Qr}
\end{abstract}

\section{Introduction}

We want to present and to illustrate a systematic scheme for the
in principle exact computation of all possible current cumulants
in Markov dynamics satisfying local detailed balance. The
algorithm is based on an identity between current and activity
fluctuations, connecting the time-antisymmetric with the
time-symmetric fluctuation sector as is typical for a dynamical
large deviation theory in nonequilibrium systems. We concentrate
here however on the mechanical aspect of the method, how it can be
seen as a modified Rayleigh-Schr\"odinger expansion with specific
computable expressions of the cumulants.  Its relevance is
therefore in reliably producing also higher-order cumulants that
can then be further analyzed for understanding the physics of some
particular model.  We will start that for an interacting particle
system with boundary driven Kawasaki dynamics.

As here we choose to emphasize the algorithm rather than its
detailed numerical implementation, we focus on relatively small
systems.  Yet we feel excused for the moment as exactly small open
systems and their nonequilibrium fluctuations have been in the
middle of attention in the last years. They are intrinsically of
relevance to nanoscale-engineering and for certain cellular and
molecular biological processes~\cite{sma,small}. Charge transport
in nano-electromechanical systems is often described in terms of
Markov evolutions, and is a subject of very active
research~\cite{rec,but,naz,nov,tom,lev}. First experiments were
limited to measuring the mean current  or its variance at most,
but now also third and higher order cumulants have become
available, providing important information on quantum
transitions~\cite{3cum,fuki}. For life processes, for instance in
molecular motors or for ion-transport through membrane channels,
one easily reaches energy scales as low as a few
$k_BT$~\cite{bio,kelvin}. Besides these cross-disciplinary
aspects, the study of all these commonly called mesoscopic systems
is important to unravel the structure of nonequilibrium
statistical mechanics itself.

Fluctuations cannot be ignored for small systems but rather carry
signatures of important physics. The computational challenge in
this case is not so much to reach large system sizes, but, for a
fixed system, to obtain the fullest possible fluctuation patterns
of the quantity of interest for long time scales. Our
results contribute to the larger effort of organizing the
computational side of the recent advances in nonequilibrium
physics, cf. \cite{dellago,posch}.  These theoretical results have
often to do with fluctuation theory, as in the Jarzynski-Crooks
relations \cite{jar,cro} or in the fluctuation theorems for the
entropy production \cite{ecm,gc,gc2,kur,lebs,poin,mn}, and going
reliably beyond Gaussian characteristics in the nonequilibrium
statistics is just a necessary but often nontrivial prerequisite.

One of the traditional approaches to nonequilibrium solid state
problems is the Keldysh formulation in terms of nonequilibrium
Green's functions~\cite{kel,crai}. Currents of any type are
obviously among the most important observables and their
fluctuations are written in the cumulants. The basic method of the
present paper comes from calculations within full counting
statistics for small quantum
systems~\cite{nov,novc,tom,gen,kin,ger,Schm,derc}. We propose yet
another derivation for classical stochastic systems
 based on a single path-distribution identity,
which allows to discuss also joint current fluctuations.
What follows can be seen as an adaptation to
the framework of Markov dynamics for the computation of joint
current cumulants in classical interacting particle systems.
In~\cite{lebs,db1,db2,jona2,jona3,derrida,dern,sei}
one finds similar treatments.
Moreover, our computational scheme aims at the same goal as in
\cite{gia,lec,rak}, but it yields exact results for small systems.
The core idea
of the method is a sufficiently simple identity, \eqref{gifu}
below, that we use in combination with expansion techniques for
eigenvalues.  The novelty in our work is then as follows:

1) We use a nonequilibrium version of the Rayleigh-Schr\"odinger
(RS) expansion to obtain a systematic cumulant expansion for the
current statistics, generalizing the approach of~\cite{novc} to
also include joint fluctuations of different currents. This is
particularly useful for the numerical evaluation of higher-order
cumulants (say from third-order on) as finite-difference
calculations generate more numerical errors. One hopes that they
are also within reach of experimental methods on real
nonequilibrium systems~\cite{3cum,fuki}. In that case they would
be invaluable tools in any attempt of reverse engineering. The RS
expansion for (in general irreversible) Markov dynamics is
computationally useful as every order in the expansion employs the
same basic information about the dynamics.
Since the generators of stochastic dynamics are not symmetric matrices, 
their set of eigenvectors might be incomplete. This is taken into account
in the solution of the problem, which involves
the use of the group pseudo-inverse of stochastic
matrices~\cite{mey1}.

2) In Section \ref{inter} we add a thermodynamic interpretation of
the numerical procedure in terms of the time-symmetric sector of
nonequilibrium fluctuations.  This lines up with the recent
introduction of the novel concept of traffic, which, roughly
speaking, measures the amount of dynamical activity in the system.
This activity counts the number of all jumps irrespectively of
their direction and hence it is symmetric under time
reversal~\cite{epl,mnw1,mnw2}.
It has also been considered in \cite{dern,app}.

3) We illustrate the procedures for a boundary driven Kawasaki
dynamics, for which an exact or analytic solution is far beyond
reach, see Section \ref{kawa}. It is interesting to
discover systematic tendencies in the role of attractive {\it
versus} repulsive potentials for the current statistics away from
equilibrium.

The paper is organized as follows: In the next section we explain
some basic identities that lead to the formulation
 of the problem as an evaluation of a certain eigenvalue.
 Section \ref{kawa} gives an explicit example where the method is
applied to a boundary driven interacting particle system. Section
\ref{inter} reflects further on the method from the point of view
of the theory of large deviations: we point out the role of a
novel concept, that of traffic, in the interpretation of the
various terms.  The paper closes with Appendices giving details
on the method and its numerical implementation.

\section{Method}

\subsection{Current fluctuations}
We suppose a continuous time Markov jump process
$(X_t)_{t\geq 0}$ on a finite state space $K$ with $M$ elements.
  The dynamics is specified by all transition rates $k(\eta,\xi)$,
from each state $\eta$ to each other $\xi\ne \eta$,
as summarized in the generator $L$,
which is a $M\times M$ matrix with elements
\begin{equation}
\begin{split}
L_{\eta\xi} &= k(\eta,\xi) \qquad\qquad \mbox{ if } \eta\neq \xi \\
L_{\eta\eta} &=  - \sum_\xi k(\eta,\xi)
\end{split}
\end{equation}
Note that the diagonal elements equal minus the respective escape
rates. We assume irreducibility in the sense that all states are
reachable from any other state in some finite time. Hence,
there is a unique stationary distribution $\rho$.
We are mostly
interested in breaking the detailed balance condition, driving the
process outside equilibrium; see Appendix \ref{mat} for some
formulation.

 The matrix $L$ generates the stochastic
evolution in the sense that
\[
\frac{d}{dt}\langle f(X_t)\rangle = \langle (Lf)(X_t)\rangle
\]
for all vectors $f: K\rightarrow \bbR$. The brackets
$\langle\cdot\rangle$ are averages both for the random (as yet
unspecified) initial conditions as over the stochastic
trajectories. The Markov process $(X_t)$ is a jump process in the
sense that the trajectories are piecewise constant (in time) with
jumps $X_t = \eta\rightarrow X_{t^+} = \xi$ from some state $\eta$ to some state
$\xi$ at random moments $t$. We consider the stationary process starting
from $\rho$.

We consider each ordered pair of connected states $b=(\eta,\xi)$
and its inverse is $-b = (\xi,\eta)$. For a given trajectory
$\omega = (X_t, 0 \leq t < T)$, over some time-span $T$ we have a
microscopic current $dJ_b(t) = +1$ when the state jumps at time
$t$ over the bond $b$, while $dJ_b(t) = -1$  when the state jumps
over the bond $-b$. The time-integrated current
\begin{equation}\label{cr}
J_b(\omega) = \int_0^T dJ_b(t)
\end{equation}
thus counts the number of net jumps over the oriented bond $b$ in
the time span $[0,T]$. Note that the dependence on $T$ in the
left-hand side of \eqref{cr} is not explicitly indicated.  If we
look at the stationary state, we should take the expectation of
\eqref{cr} and divide by $T$ to get the flux (per unit time).
 In the stationary state, the expected current over the bond $b$
and per unit time equals
$j_b = \rho(\eta)k(\eta,\xi) - \rho(\xi)k(\xi,\eta)$.

The main reason to consider all these currents like in \eqref{cr}
on the finest scale of transitions and the complexity of the full
joint fluctuations, is to be able to move to arbitrary and more
coarse-grained scales of description.  In applications, the
physical currents are all obtained by combinations of these
currents over bonds.  For example, an interesting current in a
lattice system might count the passage of particles from one given
site to another. In this case the current is rather of the form
\begin{equation}\label{cr1}
 J_B = \sum_{b\in B} J_b
\end{equation}
where $B$ then includes all $b=(\eta,\xi)$ from a state $\eta$
with a particle in the first site to a state $\xi$ where the
particle has moved to the second site (the ensemble $-B$ includes all bonds
$-b$ of the opposite transitions). This will in fact be our
main example (section~\ref{kawa}).

We can formalize that:  To keep the discussion as general as
possible, but with an eye on the actual application, we  consider
a partition of all ordered bonds (or connections) $b$'s
consisting of sets $B,B',\ldots$ for which $B, -B, B', -B',\ldots$
are mutually non-overlapping. The fully microscopic description is
recovered when each of these sets contains exactly only one bond.

We are interested in understanding and
computing the joint fluctuations of the currents $J_B$, properly
rescaled as $T\uparrow \infty$. So for example, we want to
determine the covariances
\begin{equation}\label{cov}
C_{BB'}^T = \frac 1{T} \big[\langle J_B J_{B'}\rangle - \langle
J_B \rangle\,\langle J_{B'}\rangle\big]
\end{equation}
in the large time $T$ limit for $B$ and $B'$, which corresponds to the steady
state regime. From now on, the bracket-expectations
$\langle \cdots \rangle$ refer to the mathematical expectation in the
assumed unique stationary process. Higher-order cumulants are also
important, for example to determine the deviation from Gaussian
behavior.

In general, the computation of these cumulants like in \eqref{cov}
involves detailed information about the time-autocorrelation
functions. This information is hidden in the spectrum of the
generator $L$. What we will do, amounts to extracting that
information from a systematic numerical scheme. As a further
result expressions are obtained for these cumulants in terms of
expectations of specific {\it single-time} observables under the
invariant distribution, which allows also to see relations between
the various cumulants and what governs them.

\subsection{General identity}\label{gi}
The method of computing the cumulants for the currents starts from
a general identity \eqref{gifu} that relates the current
fluctuations with fluctuations of occupation times.

We fix a set of numbers $\bmsigma=(\sigma_B)$.
The cumulant-generating function for the joint fluctuations of the
selected currents $J_B$ is then given by
\begin{equation}\label{gf}
g_T(\bmsigma) = \frac 1{T}\log\langle e^{ \sum_B \sigma_B J_B} \rangle
\end{equation}
By definition, the derivatives of
$g_T(\bmsigma)$ at $\bmsigma=0$ give us all possible
cumulants. For example the second (partial) derivatives with
respect to $\sigma_B, \sigma_{B'}$ give \eqref{cov}.  We therefore
want to obtain an expression of $g_T(\bmsigma)$ as a
Taylor-expansion in the $\sigma_B$'s, for $T\rightarrow +\infty$.

In order to do so and given the original Markov process with rates
$k(\eta,\xi)$ we now construct a new Markov process with generator
$L_\bmsigma$, where the rates
relative to bonds $b=(\eta,\xi)\in B$ and $-b=(\xi,\eta)\in -B$
are
\begin{equation}\label{modi}
\begin{split}
\ell(\eta,\xi) &= k(\eta,\xi)\,e^{\sigma_B} \\
\ell(\xi,\eta) &= k(\xi,\eta)\,e^{-\sigma_B}
\end{split}
\end{equation}
We further define the vector
\begin{equation}\label{ve}
V(\eta) = \sum_B\sum_{\xi: (\eta,\xi)\in \pm B} k(\eta,\xi)[
e^{\pm \sigma_B} - 1]
\end{equation}
where the sign in the exponent depends on whether $(\eta,\xi) =
\pm b$ for a selected bond $b$.

The generating function \eqref{gf} can be rewritten via the
identity
\begin{equation}\label{gifu}
\langle e^{\sum_B \sigma_B J_B} \rangle =
\langle e^{ \int_0^T V(X_t)\, \id t} \rangle_\bmsigma
\end{equation}
where the last average is over the Markov process with rates
$\ell(\eta,\xi)$, hence depending on $\bmsigma$.

To prove (\ref{gifu}) we note that in going between the two
averages $\langle \cdot \rangle$ and $\langle \cdot
\rangle_\bmsigma$ there is a density $e^{Q(\omega)}$,
\begin{equation}\label{rn}
\langle F(\omega)\rangle = \langle F(\omega) e^{Q(\omega)} \rangle_\bmsigma
\end{equation}
that is given by \[ Q(\omega) = \sum_t
\log\frac{k(X_t,X_{t^+})}{\ell(X_t,X_{t^+})} - \int_0^T \id
t\,\sum_\xi [k(X_t,\xi) - \ell(X_t,\xi)]
\]
where the first sum is over all jump times $t$ in $\omega$ where
the state changes $X_t \rightarrow X_{t^+}$, see for example Appendix~2
of~\cite{KL} for mathematical details. As a consequence and via
\eqref{modi},
\[ Q(\omega) = -\sum_B \sigma_B J_B + \int_0^T
V(X_t)\, \id t
\]
Substituting $F =\exp \sum_B \sigma_B J_B$ into \eqref{rn}
gives the result \eqref{gifu}.

We remark that Eq.~(\ref{gifu}) shows that the current
fluctuations can be expressed in terms of occupation time
fluctuations in a tilted path-space measure, see also in Section
\ref{inter}. It is not a new observation,
see for instance~\cite{lebs,derrida} for very related although less explicitly
stated considerations. First we continue with its
exploitation for computational purposes.

If one has only one set $B$ with $\sigma_B=\lambda\ne 0$, the
current generating function simplifies to
\begin{equation}\label{glo}
g^B_T(\lambda) = \frac 1{T}\log\langle e^{\lambda \,J_B} \rangle
\end{equation}
 The
identity \eqref{gifu} obviously remains valid, now with
\begin{equation}\label{vo}
V_B(\eta) = [ e^{\lambda} - 1]\,\sum_{(\eta,\xi)\in B} k(\eta,\xi)
+ [ e^{-\lambda} -1]\,\sum_{(\xi,\eta)\in -B} k(\xi,\eta)
\end{equation}

\subsection{Feynman-Kac formula}\label{feka}
The right-hand side of \eqref{gifu} involves the single-time
observable $V$, in contrast with a current being a double-time
function.  The $V$ can therefore be taken as a potential (diagonal
matrix) $\caV$ in the following sense: given the matrix $\caL =
L_\bmsigma + \caV$,
\begin{equation}\label{fk}
\lim_T\frac 1{T}\log \langle e^{ \int_0^T V(X_t)\, \id t} \rangle_\bmsigma
= e^{\rm{max}}_\caL
\end{equation}
where $e^{\rm{max}}_\caL$
is the largest eigenvalue (in the sense of its real part) of $\caL$.

The asymptotic formula \eqref{fk} is the limit of what is known as
the Feynman-Kac formula. For our context, one finds a proof of it
in Appendix~2 of \cite{KL}.   As a result, the current cumulants
can be read from the Taylor-expansion of the eigenvalue
$e^{\rm{max}}_\caL$ with explicitly known matrix
\[
\caL = L + \caR,\quad \caR = L_\bmsigma - L + \caV
\]
with $\caR$ having non-zero elements only for the pairs
$(\eta,\xi)$ in some $\pm B$ with $\sigma_B\ne 0$. More precisely,
for $b = (\eta,\xi) \in B$,
\begin{equation}
\begin{split}
\caR_{\eta\xi} &= k(\eta,\xi)\,[e^{\sigma_B} -1] \\
\caR_{\xi\eta} &= k(\xi,\eta)\,[e^{-\sigma_B}-1]
\end{split}
\end{equation}
Since we required that an ensemble of transitions $B$ does not
overlap with any other $B'$ or $-B'$, we can decompose the matrix
$\caR$ in a convenient sum of matrices $E_B(\sigma_B)$ and
$E_{-B}(\sigma_B)$, where each matrix $E_B$ has non-zero elements
equal to $k(\eta,\xi)\,e^{\sigma_B}$ only for $(\eta,\xi)\in B$,
and similarly each matrix $E_{-B}$ has non-zero elements equal to
$k(\xi,\eta)\,e^{-\sigma_B}$ only for $(\xi,\eta)\in -B$. Thus,
\begin{equation}\label{decR}
\caR = \sum_B\left[ E_B(\sigma_B)-E_B(0) + E_{-B}(\sigma_B)-E_{-B}(0) \right]
\end{equation}

We finally remark here that the maximum eigenvalue
$e^{\rm{max}}_\caL$ is simple, which follows again from a
Feynman-Kac formula saying that
\[ \langle e^{\int_0^T V(X_t)\, \id t}
\delta_{X_T,\xi}\rangle_{\sigma, X_0=\eta} = (e^{T \caL})_{\eta,\xi} \geq 0
\]
By the irreducibility assumption,
the left-hand side is in fact strictly
positive (for any $T > 0$ and $V$), hence the right-hand side is a
matrix with strictly positive entries. Therefore, the
Perron-Frobenius theorem implies that $\caL$ has a unique maximum
eigenvalue.  Moreover, the right and left eigenvectors of that
largest eigenvalue of $\caL$ have strictly positive coordinates.
Exactly all the same is true for the generator $L$.

\subsection{Expansion}\label{expa}
From the previous discussion it is clear that $\caR$ goes to zero with the
$\sigma_B$'s.  Moreover, there are no cross-terms containing mixed
derivatives of $\caR$ with respect to the $\sigma_B$'s.
As we recognize the cumulants of the current distribution from the
Taylor-coefficients in the eigenvalue $e^{\rm{max}}_\caL$,
it is natural to write
\[
 \caR = \sum_B\left( \sigma_B \caL^{(1)}_B  +
                  {\sigma_B^2}\,\caL^{(2)}_B +\ldots \right)
\]
over the order in the $\sigma_B$'s.
Then, for each $n=1,2,\ldots$ and $B$
\[
\caL^{(n)}_B =\frac{1}{n!}
\left[  E_B(0) + (-1)^n E_{-B}(0) \right]
\]
This means that all odd terms $n! \caL^{(n)}_B$ are
the same matrix $ E_B(0)-E_{-B}(0)$,
and that all terms $n! \caL^{(n)}_B$ with $n$ even are equal to
$ E_B(0)+E_{-B}(0)$.

From the RS perturbation expansion, see
Appendix~\ref{ap1}, we obtain the following cumulants.  It is
important to note that the computation proceeds always from the
same basic ingredients. The input consists of the stationary
distribution $\rho$ and the expression for the pseudo-inverse of
$L$, see below.  Then, {\em all} cumulants follow from an exact
numerical calculation.  More details on the algorithm are in
Appendix~\ref{ap1}.

\subsubsection{First order}
As needed, the formula for the first order cumulant corresponds to
the expectation of the current,
\begin{equation}\label{cumu1}
j_B = \lim_{T\uparrow +\infty} \frac 1{T} \mean{ J_B } = \BRA{\rho}
\caL^{(1)}_B \KET{1}
\end{equation}
where we use the Dirac notation for left and right eigenvectors:
$\BRA{\rho}$ is the density giving the steady state occupation
probabilities of the states, and $\KET{1}$ is a column vector of
$1$'s. They are the left and right eigenvectors of $L$, with
maximal (always in the sense of real part) eigenvalue $e_0=0$.

\subsubsection{Second order}
The expression for second order gives the variance
\begin{equation}\label{cumu2}
C_{BB} = \lim_{T\uparrow +\infty} C^T_{BB} = 2\; \BRA{\rho}  (
\caL^{(2)}_B - \caL^{(1)}_B \,\GI\, \caL^{(1)}_B
) \KET{1}
\end{equation}
for the current $j_B$ over bonds with field $\sigma_B$, and
the covariance \eqref{cov}
\begin{equation}\label{cov2}
\begin{split}
C_{BB'} = \lim_{T\uparrow +\infty} C^T_{BB'} =&
-\BRA{\rho} ( \caL^{(1)}_B  \,\GI\, \caL^{(1)}_{B'} ) \KET{1}\\
 & -\BRA{\rho} ( \caL^{(1)}_{B'} \,\GI\, \caL^{(1)}_B  ) \KET{1}
\end{split}
\end{equation}
if $B \ne B'$.  The matrix $\GI$
is the pseudo-inverse of the matrix $L$ in the sense of Drazin~\cite{mey1},
see Appendix~\ref{mat}.

\subsubsection{Third and fourth cumulant}
For the higher-order cumulants we restrict to the condition of a
single global current, as in \eqref{glo} and (\ref{vo}). In this case,
we have a single ensemble $B$ and
the identity \eqref{gifu} reduces to
\[
\langle e^{ \lambda J_B} \rangle =
\langle e^{ \int_0^T V_B(X_t)\,\id t} \rangle_\lambda
\]
As a result, the analogue of \eqref{fk} is verified for the matrix
$\caL = L + \caR$ with
$\caR=E_B(\lambda) - E_B(0) + E_{-B} (\lambda) - E_{-B}(0)$.
By expanding the exponential around $\lambda=0$, we write
\begin{equation}
\caR = \sum_{k=1}^{+\infty} \lambda^k \caL^{(k)}
\end{equation}
and the cumulants are obtained from the scheme outlined in the
Appendix~\ref{ap1}.

The third cumulant of the current $J_B$ over bonds $b\in B$ is
then
\begin{equation}\label{cumu3}
\begin{split}
C^{(3)} = 3!\, \BRA{\rho}\;\Big[ &
\caL^{(3)}
- j_B \caL^{(1)} \GI^2 \caL^{(1)}
+ \caL^{(1)} \GI \caL^{(1)} \GI \caL^{(1)}\\
& - \caL^{(1)} \GI \caL^{(2)}
- \caL^{(2)} \GI \caL^{(1)} \;\Big]\; \KET{1}
\end{split}
\end{equation}
and the fourth cumulant is
\begin{equation}
\begin{split}
C^{(4)} = 4!\, \BRA{\rho}\; & \Big[
\caL^{(4)} - \caL^{(2)}\GI\caL^{(2)} -\frac{C_{bb}}{2}\caL^{(1)}\GI^2\caL^{(1)}
-(j_B)^2 \caL^{(1)}\GI^3\caL^{(1)}\\
& +\caL^{(1)}\GI\caL^{(2)}\GI\caL^{(1)}
  -\caL^{(1)}\GI\caL^{(1)}\GI\caL^{(1)}\GI\caL^{(1)} \\
&  -\caL^{(3)}\GI\caL^{(1)} -\caL^{(1)}\GI\caL^{(3)} \\
& +\caL^{(2)}\GI\caL^{(1)}\GI\caL^{(1)} +\caL^{(1)}\GI\caL^{(1)}\GI\caL^{(2)} \\
& -j_B\left( \caL^{(2)}\GI^2\caL^{(1)} + \caL^{(1)}\GI^2\caL^{(2)} \right) \\
& +j_B\left(\caL^{(1)}\GI  \caL^{(1)}\GI^2\caL^{(1)}
   +\caL^{(1)}\GI^2\caL^{(1)}\GI  \caL^{(1)}\right) \;\Big]\; \KET{1}
\label{cumu4}
\end{split}
\end{equation}
Note the symmetry in the terms: when a sequence of matrices is not palindrome,
there is also its reversed one.

\section{Example}\label{kawa}
 We consider a
generalization of the symmetric exclusion process (SEP), in which,
besides via the exclusion principle, particles are also
interacting with their nearest neighbors at a finite reservoir
temperature $\beta^{-1}$. Let us consider a lattice gas on the
sites $\{1,\ldots,N\}$, where a configuration is an array of
occupations, $\eta(i) = 0,1$ for $1\le i\le N$. The state space is
thus $K = \{0,1\}^{N}$, with $M = 2^N$ different states. One can
think of particles (and holes) hopping in a narrow and small
(effectively homogeneous) channel.  The specific calculation below
has been done for a relatively small system where $N=8$.  We
comment on size-dependence of the algorithm in Appendix
\ref{apnum}

For the dynamics there are two modes of updating:  In the bulk, a
particle can jump to nearest neighbor sites. Then, the occupation
over a nearest neighbor pair of sites is exchanged. For a
transition $\eta\to\xi$ of this kind we take a rate of
the form
\begin{equation}\label{cij}
k(\eta,\xi) = \exp \Bigl[ -\frac{\beta}{2}(H(\xi)-H(\eta))\Bigr]
\end{equation}
where $H$ is the  energy function
\begin{equation}\label{ham}
H(\eta) = -\epsilon \sum_{i=1}^{N-1} \eta(i)\eta(i+1)
\end{equation}
for some parameter $\epsilon$. Thus, only pairs of
particles occupying nearest neighbor sites have an energetic contribution.

At the boundaries one has the second kind of updating. At
site $i=1$ particles can be exchanged with an external reservoir
having chemical potential $\alpha/\beta$. In the case of a
particle leaving the system ($\eta(0)=1 \to \xi(0)=0$) the rate is
given by
\begin{equation}\label{ci}
k(\eta,\xi) = \exp\Bigl[
-\frac{\alpha}{2}-\frac{\beta}{2} (H(\xi) - H(\eta))\Bigr]
\end{equation}
while a particle enters into the system at site $i=1$ with rate
\begin{equation}\label{ci+}
k(\eta,\xi) = \exp
\Bigl[\frac{\alpha}{2}-\frac{\beta}{2} (H(\xi) - H(\eta))\Bigr]
\end{equation}
We focus on the time-integrated current $J$ passing through the
site $i=1$, which increases by $1$ every time a particle enters
there from the reservoir and decreases by $1$ every time a
particle leaves the system from there. As explained in previous
sections, this is the sum of all microscopic currents $J_b$ over
bonds $b$ connecting a state $\eta$ with $\eta(1)=0$ to another
state $\xi$ with $\xi=\eta$ on all sites except $\xi(1)=1$. At the
other boundary site $i=N$ a similar structure may be imposed, with
chemical potential $\alpha'/\beta$. The time-integrated current
$J'$ however is defined there with the opposite convention, i.e.,
one adds one to $J'$ when a particle leaves the system. With this
convention, both currents $j=J/T$ and $j'=J'/T$ have the same
asymptotic value for a time span $T\to \infty$, as they both flow
from left to right.

The model is a boundary driven Kawasaki dynamics, reducing  to
boundary driven SEP for $\beta=0$.  This infinite temperature case
is completely solved concerning current fluctuations in
\cite{derex,bd3}. If $\alpha=\alpha'$, then it is easily checked
that the model satisfies the condition of detailed balance with
respect to the grand-canonical distribution for energy \eqref{ham}
and chemical potential $\alpha/\beta$. If however $\alpha\neq
\alpha'$ then the system is driven out of equilibrium: the
difference in effective chemical potential between reservoirs
generates a particle current through the system. It is a
fluctuating current and we study here its cumulants. For other
models, similar questions have been addressed for example in
\cite{zero,weak}. Studies on the density fluctuations for the
boundary driven exclusion process are in \cite{der,jona1}.

For simplicity, we set $\alpha'=0$ and drive the system by varying
only $\alpha$ and $\beta$. The case $\alpha>0$ thus corresponds to
a reservoir that pushes particles from the left into the system,
forcing a positive stationary current $j$. The case $\alpha<0$
instead corresponds to a left reservoir that tends to remove
particles. As we will see, the two situations are definitely not
the mirror image of each other (unless $\beta=0$).

Since the product $\beta \epsilon$ is what matters in the transition
rates, we simply set $\epsilon=1$ and we use the possibility
$\beta<0$ for characterizing repulsive potentials. Particles instead
attract each other for $\beta>0$. Particle interactions very much
complicate the model which is no longer analytically tractable. We
use the above formalism to evaluate the current cumulants for
different parameter values. Interestingly, interactions induce
qualitatively novel behavior for the current statistics.

\subsection{Mean current}

\begin{figure*}[!tb]
\begin{center}
\includegraphics[angle=0,width=11cm]{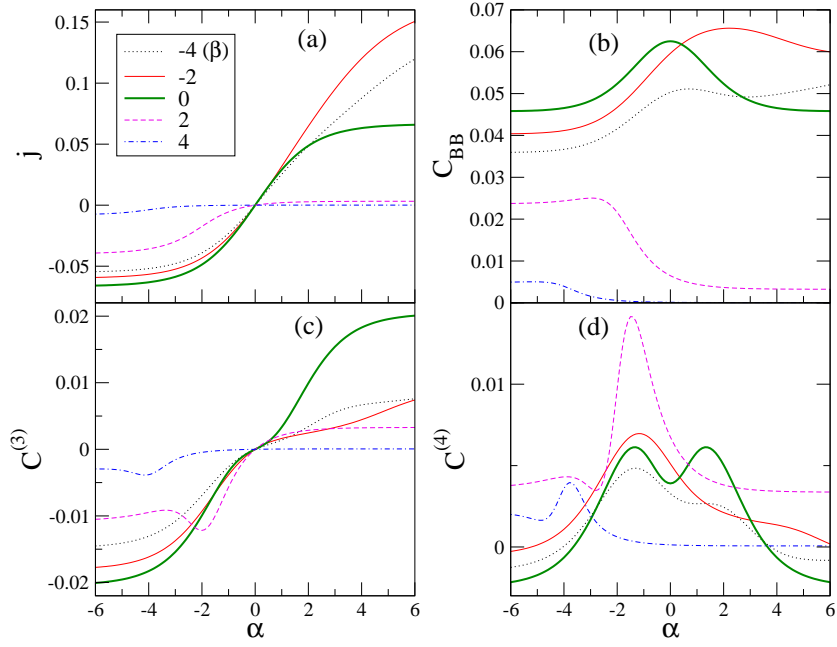}
\end{center}
\caption{(Color online)
First four cumulants of the current distribution as a function
of $\alpha$, for 5 values of the interaction ``strength'' $\beta$
(see the legend).
Note that for $\beta=0$ (SEP) odd cumulants are antisymmetric functions of the
driving, while even cumulants are symmetric functions. This is due to a
particle-hole symmetry, which is lost for interacting particles.
\label{fig:vsalpha}}
\end{figure*}

\begin{figure*}[!tb]
\begin{center}
\includegraphics[angle=0,width=11cm]{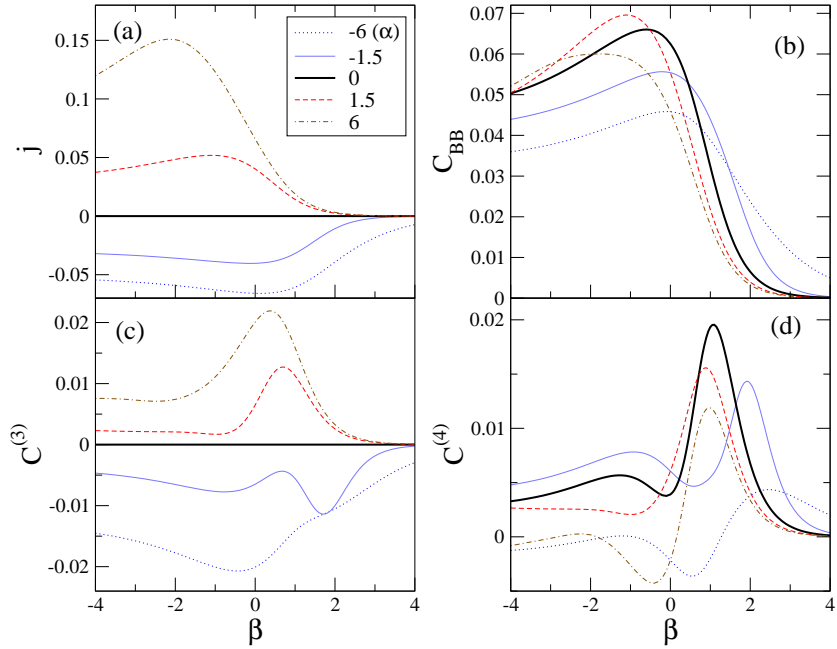}
\end{center}
\caption{(Color online) 
First four cumulants of the current distribution as a function
of $\beta$, for 5 values of the driving $\alpha$ (see the legend).
Odd cumulants are identically zero when the system is in equilibrium
($\alpha=0$).
\label{fig:vsbeta}}
\end{figure*}

The mean current $j$ as a function of  $\alpha$ and for several $\beta$'s
is shown in figure~\ref{fig:vsalpha}(a). For a given $\beta$, $j$ increases
with $\alpha$, linearly around $\alpha=0$, as expected close
to equilibrium. For each $\alpha>0$ the mean
current is maximal for a repulsive interaction ($\beta<0$), see the two
examples in figure~\ref{fig:vsbeta}(a)).
In general, the mean current $j$ is not antisymmetric with respect to
$\alpha$, and its value in $\alpha$ can be very different from
$-j$ in $-\alpha$.

For $\beta\to -\infty$ the problem can be mapped into the dynamics of
non-interacting dimers ``$(0,1)$'' and of $0$'s.
In this limit the system is somewhat like a SEP with $1$'s replaced by
dimers, and one thus expects a finite mean current.
On the other hand, for $\beta\to+\infty$, particles stick together
and it becomes more and more difficult for a hole (vacancy) to get
in, to reach the bulk and finally to reach the other boundary of the channel.
The hole essentially performs a random walk with an open left
boundary before eventually reaching the system at the right
boundary. If more than one hole enters into the system, there is a good chance
that holes stick, further reducing the energy of the
system,  and their own mobility and $j$
as well. Thus, for $\beta\to\infty$ we expect $j\to 0$.
These scenarios are qualitatively well confirmed in figure~\ref{fig:vsbeta}(a).

For $\beta=0$ one has the well-studied driven SEP. In this case
the current is antisymmetric with $\alpha$, like all odd
cumulants, because of a corresponding particle/hole symmetry.

\subsection{Variance}

The second cumulant of the current distribution is its variance.
For $\beta=0$ the variance is symmetric with $\alpha$ (as every other even
cumulant), while for
all other cases it displays a non-trivial dependence on $\alpha$ and
$\beta$,  see figures~\ref{fig:vsalpha}(b) and~\ref{fig:vsbeta}(b).
For $e^\beta\gg 1$ (see $\beta=4$ in figure~\ref{fig:vsalpha}(b)) the variance,
as the current, can reach very small values, confirming the scenario
proposed above.

\subsection{Third and fourth cumulants}

As for the current, for $\beta=0$ (SEP) the third cumulant of the
current is antisymmetric with $\alpha$, see
figure~\ref{fig:vsalpha}(c). In general, however, it is a
complicated function of $\alpha$ and $\beta$, as evidenced  by
figure~\ref{fig:vsbeta}(c). For example, in contrast with the mean
$j$, it can be a non-monotonous function of $\alpha$. Similar
arguments apply to the fourth cumulant, see
figure~\ref{fig:vsalpha}(d) and figure~\ref{fig:vsbeta}(d). The
third and the fourth cumulant also appear going to zero for
$e^\beta\gg 1$.

\section{Traffic}\label{inter}

Systematic perturbation techniques better be accompanied by a
larger theoretical understanding.  A major step in the analysis of
the problem at hand that proceeds the numerical algorithm is
contained in the simple path-distribution identity (\ref{gifu}).
On the left-hand side, this identity involves an average over
paths $\omega$ in the original process, with probabilities
$\Prob(\omega)$. The respective probabilities in the tilted space
(with rates \eqref{modi}) can be written as
$\Prob_\bmsigma(\omega) = e^{-Q(\omega)} \Prob(\omega)$, with
relative path-space action $Q$. Since on the left-hand side of
identity  (\ref{gifu}) we have a current generating function, a
time-antisymmetric quantity is involved. On the other hand, on the
right-hand side of (\ref{gifu})
 only a potential $V$ appears,
i.e., a quantity depending only on the states and thus insensitive
to time-reversal. Hence, the choice of the tilted Markov process
is exactly such that the change in the time-antisymmetric part of
the path-space action equals the appropriate sum over currents.
This is why the exponent in the right-hand side of \eqref{gifu}
contains a time-symmetric function only.

Such considerations are typical of the Lagrangian approach to
nonequilibrium statistical mechanics as pioneered by Onsager and
Machlup, \cite{ons}.  Here however we are not even close to
equilibrium.  We thus move on a somewhat generalized formalism
that remains quite simple for finite state space Markov processes.
Nevertheless the structure of time-symmetric versus
time-antisymmetric fluctuations is possibly important for
nonequilibrium thermodynamics, if only to identify the relevant
thermodynamic potentials, cf. \cite{epl,mnw1,mnw2}.  Such an
identification proceeds via a dynamical fluctuation theory, in
which we next situate the main identity \eqref{gifu}.

In order to rewrite (\ref{gifu}) in another convenient form,
we define occupation times as
\[
\mu_\eta(\omega) = \frac 1{T} \int_0^T\, \id t\; \delta_{X_t,\eta}
\]
that the path $\omega=(X_t, 0<t\leq T)$ spends in state $\eta$.
Then, the exponent in the right-hand side of \eqref{gifu} equals $
\int_0^T V(X_t)\, \id t = T \sum_\eta V(\eta) \mu_\eta(\omega) $ or,
\begin{equation}\label{29}
 \langle e^{\sum_B \sigma_B J_B}\rangle =
\langle e^{ T \sum_\eta V(\eta) \mu_\eta } \rangle_\bmsigma
\end{equation}
The current statistics is therefore obtained when one knows the
large deviation rate function $I^\bmsigma$ for the occupation times,
\[
\Prob_\bmsigma[\mu_\eta \approx p(\eta), \forall \eta] \sim
e^{-T I^\bmsigma(p)},\quad T\uparrow +\infty
\]
for the modified dynamics \eqref{modi}: \begin{equation}\label{va}
\lim_{T} \frac 1{T} \log \langle e^{\sum_B \sigma_B J_B}\rangle =
\sup_{\mu} (\mu \cdot V - I^\bmsigma(\mu)) \end{equation}
We have
in mind here the application of the theory of large deviations as
pioneered in \cite{DV} for Markov processes.

In that last variational expression \eqref{va}, the potential $V$
also depends on $\bmsigma$.  Let us introduce the antisymmetric
form $\sigma(\eta,\xi) = \sigma_B$ for $(\eta,\xi) = b$ and
$\sigma(\xi,\eta) = - \sigma(\eta,\xi)$.  Then, the change
\eqref{modi} from the original rates $k(\eta,\xi)$ to the new
rates $\ell(\eta,\xi)$ adds a further driving (in the spirit of
local detailed balance): from \eqref{ve}, the term
\begin{equation}\label{tra}
\begin{split}
 \mu \cdot V &= \sum_\eta \mu(\eta) \,V(\eta) \\
 &= \sum_\eta \mu(\eta) \, \sum_\xi k(\eta,\xi)[ e^{\sigma(\eta,\xi)} -
 1] \\
&= \frac 1{2}\sum_{\eta,\xi}[\tau_{\mu,\ell}(\eta,\xi) -
\tau_{\mu,k}(\eta,\xi)]
\end{split}
\end{equation}
is an expected excess traffic, defined for rates $k$ as
\[
\tau_{\mu,k}(\eta,\xi) = \mu(\eta) \, k(\eta,\xi) + \mu(\xi) \,
k(\xi,\eta)
\]
and similarly for rates $\ell$, see \cite{epl,mnw1,mnw2}.
The traffic expresses a
time-symmetric kind of dynamical activity over the bond
$b=(\eta,\xi)$.  In fact, all cumulants of the expansion in
Section \ref{expa} contain the term
\begin{equation*}
n!\,\BRA{\rho}\; \caL^{(n)}_B\; \KET{1} =
\begin{cases}
j_B & \text{for $n$  odd} \\
\tau_B & \text{for $n$ even}
\end{cases}
\end{equation*}
with expected current over bonds $b\in B$
\[
j_B
=  \BRA{\rho} [E_B(0)-E_{-B}(0)]\KET{1}
\]
and with corresponding expected traffic
\[
\tau_B
=\BRA{\rho} [E_B(0)+E_{-B}(0)] \KET{1}
\]
For instance, the first term on
the right-hand-side of \eqref{cumu2} is the stationary traffic
$\tau_B$, while the second term
can be interpreted as a zero-frequency autocorrelation function.

We stress that the traffic $\tau_B$ is symmetric under the
exchange $\eta\leftrightarrow\xi$, while the current $j_B$ is
antisymmetric. In other words, the traffic adds a time-symmetric
aspect to the evaluation of the dynamical activity. Finally, note
that the following identity holds,
\[
\BRA{\rho}  \caR \KET{1} =
\sum_B \left[\tau_B \,(\cosh \sigma_B -1) + j_B \,\sinh \sigma_B\right]
\]

\section*{Acknowledgements}
M.~B.~acknowledges financial support from K.~U.~Leuven grant OT/07/034A.
C.~M.~benefits from the Belgian Interuniversity Attraction Poles Programme
P6/02.
K.~N.~thanks Tom\'a\v{s} Novotn\'y for fruitful discussions
and acknowledges the support from the project AVOZ10100520 
in the Academy of Sciences of the Czech Republic and from the 
Grant Agency of the Czech Republic (Grant no.~202/07/J051).

\appendix

\section{Markov generator and its normality}\label{mat}

The operator $L$ that generates the Markov dynamics is a $M\times
M$ matrix, and its spectral properties appear in the expansion for
the cumulants (see more in the next appendix). It is important to
realize some important changes with respect to equilibrium. For an
equilibrium process with reversible distribution $\rho > 0$ there
is detailed balance,
\[ \rho(\eta)\,L_{\eta\xi} = \rho(\xi)\,L_{\xi\eta}
\]
Equivalently, the matrix \[ H_{\eta\xi} =
\sqrt{\rho(\eta)}\,L_{\eta\xi}\,\frac 1{\sqrt{\rho(\xi)}} \] is
then symmetric and hence diagonalizable with a complete
orthonormal set of eigenvectors. The matrix $H$ is obtained from
$L$ via a similarity transformation $H=Q^{-1}LQ$ with here, and
that is essential, a diagonal similarity matrix $Q$. In other
words, we easily find a scalar product for which the eigenvectors
of a detailed balance generator are orthonormal. All that need not
be possible for nonequilibrium processes.

A central notion here is that of normality:  a matrix is normal if
and only if it commutes with its adjoint if and only if it has a
complete orthonormal set of eigenvectors. Detailed balance generators are
similar with diagonal $Q$ to normal matrices while nonequilibrium
processes have generators that need not be similar to normal
matrices at all. When such a generator is similar to a normal
matrix, then it is diagonalizable and we can work with a
bi-orthogonal family of left/right eigenvectors. The following
example illustrates some of these points.

Take the fully symmetric $3$-state Markov process, i.e. with all
rates equal to 1, and perturb it obtaining the generator\[ \left(
\begin{array}{ccc}
   -2-f+g  & 1+f &   1-g \\
    1-f  &  -2+f-h & 1+h\\
    1+g   &  1-h  & -2-g+h
 \end{array} \right)\]
 in the region $|f|,|g|,|h| < 1$. The condition of detailed balance
is satisfied on the surface $f + g + h + f g h = 0$. The nature of
the spectrum depends on the sign of $D = f g + f h + g h$:  if $D
< 0$, then the generator is diagonalizable and has real
eigenvalues; if $D = 0$ and at least one of the $f,g$ or $h$ is
non-zero, then the matrix is not diagonalizable; if $D > 0$ then
it is diagonalizable with complex eigenvalues. In particular, all
three cases occur arbitrarily close to the reference equilibrium
$f  = g = h = 0$.

One consequence of the above facts directly concerns the expansion
and calculation of the cumulants following the scheme of Appendix
\ref{ap1}.  We cannot simply rely on making use of some of the standard
tools of quantum mechanical calculation, like decomposition in an
orthonormal basis.  An important example concerns the calculation
of the pseudo-inverse as in \eqref{cumu2}--\eqref{cov2}. When we
still have a decomposition of the unity in terms of left/right
eigenvectors, then the pseudo-inverse $\GI$ can be obtained from
\[
\GI =
(\caI-P)\frac 1{L}(\caI-P) = \sum_{v=1}^{M-1} \KET{w_v^{(0)}} \frac
1{e_v^{(0)}} \BRA{\rho_v^{(0)}}
\]
with $\BRA{\rho_v^{(0)}}$ and $\KET{w_v^{(0)}}$ left and right eigenvectors
of $L$ with eigenvalue $e_v^{(0)} < 0$, and where
\[
P\equiv \KET{1}\BRA{\rho}
\]
is the projection on the vector space of constant functions (therefore
$\caI-P$ is the projection on the space orthogonal to them).
In our general case, we employ the
group inverse, a special case of Drazin inverse, see \cite{mey1}.
Its role for the computational theory of Markov processes has been
advocated in \cite{mey2}.
The group inverse of $L$ is the unique solution $\GI$ of the
equation
\[
L\,\GI \,L = L,\quad \GI\,L\,\GI = \GI,\quad L\,\GI = \GI\,
L
\]
As will appear in the next section, and as visible already in
\eqref{cumu2}--\eqref{cov2} and \eqref{cumu3}--\eqref{cumu4},
that pseudo-inverse appears in the cumulant expansion.

A final important difference between symmetric versus
non-symmetric matrices (up to a diagonal similarity
transformation) concerns the application of a variational
principle to characterize the maximal eigenvalue. For example, in
quantum mechanics one usefully employs the Ritz variational
principle for Hamiltonians (Hermitian matrices) and for finding
the ground state energy.  We are not aware of an extension of that
Ritz variational method or of a more general minimax principle to
non-Hermitian matrices.  The only variational characterization that
seems to remain goes undirectly via the relation of the largest
eigenvalue to a suitable generating function, like in (\ref{fk}), which itself
obtains a variational expression in terms of a large deviation rate
function, like in formula \eqref{va}.

\section{Rayleigh--Schr\"odinger expansion: the algorithm}\label{ap1}

We give a review of the expansion that is used to compute
the leading orders in the maximal eigenvalue.  We refer to pages
74--81 in the book of Kato~\cite{Kato}, for full details and
for a rigorous treatment.

The RS expansion finds its origins
in quantum mechanical problems of time-independent perturbation
theory~\cite{RSP1,RSP2,RSP3}. In contrast with the situation in
quantum mechanics or with the case of detailed balance, we have in
general no scalar product for which $L$ has an orthonormal basis
of eigenvectors.  In many cases in nonequilibrium, we do have a
bi-orthogonal family of $M$ eigenvectors (instead of the
orthonormal family in quantum mechanics) but it also
happens that the generator is not diagonalizable and that we have
no appropriate basis to express most easily the expansion.
Fortunately, all that is not necessary and the expansion can
proceed in a more general way.  One simplifying feature is that
the maximal eigenvalue that we need to compute
is simple, as shown in Section \ref{feka}.
For the purpose of the present Appendix,
we also make the simplification
that only one $\sigma_B=\sigma\ne 0$.

The starting point is the $M \times M$ matrix $L + \caR$ that we
write in expansion
\begin{equation}\label{rex}
\caL = L + \caR = \sum_{k=0}^{\infty} \sigma^k \caL^{(k)},\qquad
\caL^{(0)} = L
\end{equation}
The unperturbed generator $L$ has a resolvent $r(\kappa) =
(L-\kappa)^{-1}$ with Laurent series around $\kappa=0$ given by
\begin{equation}\label{reso}
\frac 1{L-\kappa} = -\frac 1{\kappa}\,P +
\sum_{m=0}\kappa^m{\GI}^{m+1}
\end{equation}
for the projection $P= \KET{1}\;\BRA{\rho}$
on the eigenspace of eigenvalue zero,
and $\GI$ the pseudo-inverse in the sense
of Drazin as we had in the previous section.

The resolvent for $\caL$ is
\[
r(\sigma,\kappa) = \frac 1{\caL - \kappa}
\]
defined for all $\kappa$ not equal to any of the eigenvalues of
$\caL$.  It can be written as a power series in $\sigma$ around
\eqref{reso}:
\begin{equation}\label{reso1}
r(\sigma,\kappa) = r(\kappa) +
\sum_{n=1}^{+\infty}\sigma^n\,r^{(n)}(\kappa)
\end{equation}
with \[
 r^{(n)}(\kappa) = \sum_{\nu_1+\ldots+\nu_p=n}(-1)^pr(\kappa)
\caL^{(\nu_1)}r(\kappa)\caL^{(\nu_2)}\ldots
\caL^{(\nu_p)}r(\kappa)
\]
where the sum is over all $1\leq p\leq n, \nu_i\geq 1$.  On the
other hand, by Cauchy's residue theorem
\begin{equation}\label{cau}
e(\sigma) = -\frac{1}{2\pi i} \Tr\oint_\Gamma \kappa
\,r(\sigma,\kappa)\,\id \kappa
\end{equation}
for a circle $\Gamma$ enclosing zero but no other eigenvalues of
$L$.  Upon substituting \eqref{reso1} into \eqref{cau} we obtain
\begin{equation}\label{sub1}
e(\sigma) = -\frac{1}{2\pi i} \Tr\oint_\Gamma \kappa
r(\kappa)\sum_{p=1}^{+\infty}\big[-\caR\,r(\kappa)\big]^p\,\id
\kappa
\end{equation}
where $\caR$ of course depends on $\sigma$.
Since $\frac{\id}{\id\kappa}r(\kappa) = r(\kappa)^2$, we have
\[
\begin{split}
\frac{\id}{\id\kappa}\big[\caR\,r(\kappa)\big]^p=&
\caR\,r(\kappa)\ldots \caR\,r(\kappa)\caR\,r(\kappa)^2 \\
& +
\caR\,r(\kappa)\ldots \caR\,r(\kappa)^2\caR\,r(\kappa)  \\
& + \ldots +
\caR\,r(\kappa)^2\ldots\caR\,r(\kappa) \caR\,r(\kappa)
\end{split}
\]
Observe now that the trace and the integration commute so that
\eqref{sub1} becomes
\[
e(\sigma) = -\frac{1}{2\pi i} \Tr\oint_\Gamma \kappa
\sum_{p=1}^{+\infty}\frac
1{p}\frac{\id}{\id\kappa}\big[-\caR\,r(\kappa)\big]^p\,\id
\kappa
 \]
and after integration by parts
\[
e(\sigma) = \frac{1}{2\pi i} \Tr\oint_\Gamma
\sum_{p=1}^{+\infty}\frac 1{p}\big[-\caR\,r(\kappa)\big]^p\,\id
\kappa
 \]
or
\begin{equation}\label{sub2}
e(\sigma) = -\frac{1}{2\pi i} \Tr\oint_\Gamma
\log\big[1+\caR\,r(\kappa)\big]\,\id \kappa
 \end{equation}
Expanding the logarithm with \eqref{rex} makes the expansion of
the maximal eigenvalue
\begin{equation}
e(\sigma) =
\sum_{n=1}^{+\infty}\sigma^n e^{(n)}=
\sum_{n=1}^{+\infty}\sigma^n \frac{C^{(n)}}{n!}
 \end{equation}
 for
\begin{equation}\label{eex}
e^{(n)} = \frac{1}{2\pi i} \Tr
\sum_{\nu_1+\ldots+\nu_p=n}\frac{(-1)^p}{p}\oint_\Gamma
\caL^{(\nu_1)}r(\kappa)
 \ldots \caL^{(\nu_p)}r(\kappa)\,\id\kappa
 \end{equation}
We finally substitute the series \eqref{reso} and perform the
integral again with the residue theorem to get the result
\begin{equation}\label{eq:en}
e^{(n)} = \sum_{p=1}^{n}
\frac{(-1)^p}{p}\sum_{\substack{\nu_1+\ldots+\nu_p=n \\ k_1+\ldots+k_p=p-1}}\Tr
\caL^{(\nu_1)}S^{(k_1)}\ldots\caL^{(\nu_p)}S^{(k_p)}
 \end{equation}
where $S^{(0)} = -P$ and $S^{(k)} = \GI^k$.
The last formula can be written more explicitly obtaining the
different orders $C^{(n)} = n!\, e^{(n)}$ as in Section \ref{expa}.
As an example, let us show how to compute the cumulant of order $n=2$.
The possible cases in the
first sum of~(\ref{eq:en}) are then $p=1$ and $p=2$.

For $p=1$ the second sum can only have $\nu_1=2$ and $k_1=0$, hence
the contribution is $- \Tr \caL^{(2)} S^{(0)}$. It is convenient to use the
cyclic property of the trace operator $\Tr A B = \Tr B A$, and the
definition $S^{(0)}=-P$ to rewrite the term as $\Tr P  \caL^{(2)}$.
In general, given a set of left eigenvectors $\BRA{\rho_\ell}$ and
right eigenvectors $\KET{w_\ell}$, for the trace one has
$\Tr A = \sum_\ell \BRA{\rho_\ell} A \KET{w_\ell}$. Here, the projection $P$
on the $0$-th eigenvectors ($\BRA{\rho}$ and $\KET{1}$)
simplifies this term to $\BRA{\rho} \caL^{(2)} \KET{1}$.

The only combinations of two numbers summing up to $p=2$ is
$(\nu_1=1,\nu_2=1)$, while there are two choices $(k_1=1,k_2=0)$ and
$(k_1=0,k_2=1)$ summing up to $p-1=1$.
The former case corresponds to
${1\over 2} \Tr \caL^{(1)} S^{(1)}\caL^{(1)} S^{(0)}$
$= {1\over 2} \Tr \caL^{(1)} G \caL^{(1)} (-P)$
$= {1\over 2} \Tr -P \caL^{(1)} G \caL^{(1)}$, which is equal,
according to previous arguments, to
$-{1\over 2} \BRA{\rho} \caL^{(1)} G \caL^{(1)}\KET{1}$.
The same is true for the second term
${1\over 2} \Tr \caL^{(1)} S^{(0)}\caL^{(1)} S^{(1)}$,
and their sum cancels the factor $1/2$.
Hence, overall one has the second cumulant given in Eq.~(\ref{cumu2}).

\section{Numerical scheme}\label{apnum}

We have shown that all that is required for the computation of
cumulants, regardless of their order, is the information on the
stationary distribution $\BRA{\rho}$ and on the group inverse of the
generator, i.e.\ the matrix $\GI$. An efficient computation of
$\GI$ thus enables really making use of our formulas for the
cumulants, like in Eq's.(\ref{cumu1})-(\ref{cov2}), (\ref{cumu3}),
and (\ref{cumu4}). Given $G$ and $\rho$, each cumulant is computed
just by some matrix multiplications. The estimate of the group
inverse of a generator $L$ is discussed in section 5
of~\cite{mey2} and in~\cite{sonin}. In the computations carried
out in this work, it turned out that
\[
\GI = P + (L-P)^{-1}
\]
was the most stable way of computing $\GI$ for all parameter values.
This formula derives most conveniently by using the 
properties of the so called fundamental matrix, see~\cite{heyman}.

However, for systems with a large number of degrees of freedom, it is
rarely a good idea to directly invert matrices. Fortunately, it is also
not necessary here. Note that any vector $\KET{z} = G\KET{y}$ 
coincides with the (unique) solution of the equation 
$L\KET{z} = (\caI-P)\KET{y}$ constrained by the
condition $\BRAKET{\rho}{z} = 0$, as immediately follows from observing that  
$LG = GL = \caI-P$ and $\BRA{\rho} G = 0$. 
Hence, objects like $G L^{(1)} G L^{(1)}\KET{1}$ 
or $G^2 L^{(2)}\KET{1}$ can most conveniently be determined by solving a linear
system of $M$ equations with subsequently updated right-hand side. The
number of such linear problems is fixed by the order of cumulants to be
computed. This formulation also invites an application of fast iterative
methods and various schemes to store sparse matrices in the memory, which
enable to remarkably increase the system size. 

The second basic
ingredient of the proposed algorithm is the computation of the stationary
distribution $\BRA{\rho}$, for which one can choose among the available
algorithms on the market. A possibility is to implement an Arnoldi
scheme: the iteration of $\BRA{\rho_{i+1}}\leftarrow
\BRA{\rho_{i}} (L + c \caI)$ converges to the eigenvector of $(L +
c \caI)$ with largest modulus, which coincides with $\BRA{\rho}$
if the real constant $c>0$ is larger than the modulus of all
eigenvalues of $L$.

Let us finally stress that the estimates of cumulants obtained in
this paper, besides having their own theoretical interest, have
the advantage of avoiding the use of finite differences, in this
case of eigenvalues of $L_\bmsigma$ obtained at different values
of the parameters $\bmsigma$.  Like it is convenient to estimate the
specific heat of a system from the variance of the energy
distribution rather than from finite differences of the energy at
different temperatures, 
we avoid the calculation of derivatives from finite differences, 
also because they usually hide dangerous dependencies
on parameter step-sizes and the numerical instability connected
with this. The latter is expected to be particularly problematic
for cumulants of higher order.


\end{document}